\documentclass[12pt,preprint]{aastex}
\usepackage{amsmath}

\newcommand{\Swift}{{\it Swift}}
\newcommand{\Suzaku}{{\it Suzaku}}

\newcommand{\fermi}{{\it Fermi }}

\slugcomment{}
\shorttitle{Energetic Fermi/LAT GRB100414A}
\shortauthors{Urata  et al.}

\begin{document}

\title{ENERGETIC FERMI/LAT GRB100414A: ENERGETIC AND CORRELATIONS}

\author{
Yuji~\textsc{Urata}\altaffilmark{1}, 
Kuiyun~\textsc{Huang}\altaffilmark{2},
Kazutaka~\textsc{Yamaoka}\altaffilmark{3},
Patrick~P.~\textsc{Tsai}\altaffilmark{1}, 
and 
Makoto~S.~\textsc{Tashiro}\altaffilmark{4}
}

\altaffiltext{1}{Institute of Astronomy, National Central University, Chung-Li 32054, Taiwan, urata@astro.ncu.edu.tw}
\altaffiltext{2}{Academia Sinica Institute of Astronomy and Astrophysics, Taipei 106, Taiwan}
\altaffiltext{3}{Department of Physics and Mathematics, Aoyama Gakuin University, 5-10-1, Fuchinobe, Sayamihara 229-8558, Japan}
\altaffiltext{4}{Department of Physics, Saitama University, Shimo-Okubo, Saitama, 338-8570, Japan}

\begin{abstract}

  This study presents multi-wavelength observational results for
  energetic GRB100414A with GeV photons.  The prompt spectral fitting
  using \Suzaku/WAM data yielded spectral peak energies of $E^{\rm
    src}_{\rm peak}$ of $1458.7^{+132.6}_{-106.6}$ keV and $E_{\rm
    iso}$ of $34.5^{+2.0}_{-1.8}\times10^{52}$ erg with $z=1.368$. The
  optical afterglow light curves between 3 and 7 days were effectively
  fitted according to a simple power law with a temporal index of
  $\alpha=-2.6\pm0.1$. The joint light curve with earlier \Swift/UVOT
  observations yields a temporal break at 2.3$\pm$0.2 days. This was
  the first \fermi/LAT detected event that demonstrated the clear
  temporal break in the optical afterglow. The jet opening angle
  derived from this temporal break was $5^{\circ}.8$, consistent with
  those of other well-observed long gamma-ray bursts (GRBs).  The
  multi-wavelength analyses in this study showed that GRB100414A
  follows ${E^{\rm src}_{\rm peak}-E_{\rm iso}}$ and ${E^{\rm
      src}_{\rm peak}-E_{\rm \gamma}}$ correlations. The late
  afterglow revealed a flatter evolution with significant excesses at
  27.2 days. The most straightforward explanation for the excess is
  that GRB100414A was accompanied by a contemporaneous supernova. The
  model light curve based on other GRB$-$SN events is marginally
  consistent with that of the observed lightcurve.

\end{abstract}

\keywords{Gamma-ray burst: individual}

\section{Introduction}

The Large Area Telescope (LAT) on board \fermi {\it Gamma-ray Space Telescope}
has opened a new window for observing gamma-ray bursts (GRBs) from the
range of 20 MeV to 300 GeV
(e.g., \citet{090902b, 080916c,090926a,090510}).  
Characterizing highly 
energetic events involving GeV photons is one of the most
important issues in determining their origin.
Similar to other GRBs, prompt spectral analyses and optical monitoring
observations are critical for addressing this topic
(e.g. \citet{080916c-opt}). Optical monitoring observations enable the
establishment of constraints on the jet opening angle and offer
insights into the possible progenitor.  Although spectral
confirmations are generally required, the bumps in the late optical
afterglow light curves are usually interpreted as a superimposition of a
supernova component associated with GRB.  Multi-band analysis can
reveal the existence of tight correlations linking several properties
of GRB, namely the spectral peak energy, total radiated energy,
and the afterglow break time \citep{amati, g07}. The discovery of
these correlations can provide critical insights into crucial areas of
GRB physics that are not yet completely understood.  Although some
outliers were found in the tight correlations $E^{\rm src}_{\rm peak}$-$E_{\rm \gamma}$ (e.g., GRB071010B; Urata et al. 2009, GRB050904;
Sugita et al. 2009), the $E^{\rm src}_{\rm peak}$-$E_{\rm iso}$ might
yet provide information on the characteristics of different subclasses
of GRBs \citep{amati06}.

This Letter presents the spectral analysis of the prompt gamma-ray and
systematic optical follow-up results for GRB100414A.  This event was
detected and localized using \fermi/LAT\citep{lat}.  More than 20
photons above 100 MeV were observed within 300 s, and the highest
energy photon was a 4 GeV event observed 40 s after the burst
\citep{lat}. The \fermi/GBM\citep{gbm}, Konus/{\it WIND}\citep{konus},
and \Suzaku/WAM\citep{wam-obs} observations also detected this main
burst, while the Inter Planetary Network further confirmed the
localization \citep{ipn}. \citet{xrt} improved the GRB position by
identifying the X-ray afterglow using \Swift/XRT.  The optical
afterglow was discovered by \citet{redshift} using Gemini/GMOS. The
redshift was determined to be $z=1.368$ according to a series of metal
absorption features \citep{redshift}.

We use the power-law representation of flux density, $f_{\nu}(t) \propto
t^{-\alpha}\nu^{-\beta}$, where $\alpha$ and $\beta$ are the temporal
and spectral indices, respectively.  All errors are quoted at
1$\sigma$ confidence level in this Letter.

\section{Observations and Results}

\subsection{Prompt emission}

\begin{figure}
\epsscale{.80}
\plotone{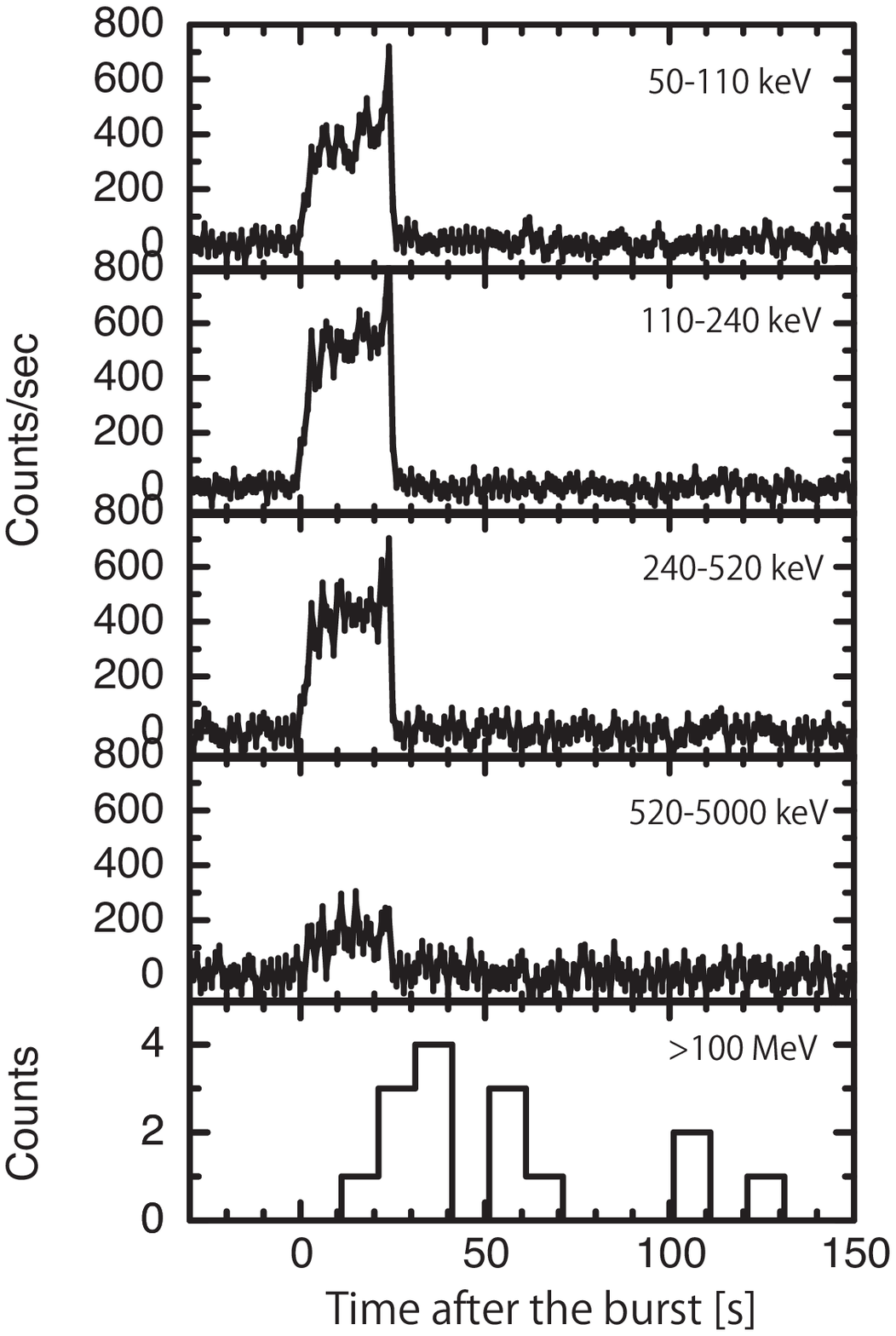}
\caption{Prompt X-ray and gamma-ray light curves of GRB100414A, as 
  observed using \Suzaku/WAM and {\it Fermi}/LAT. \label{wamlc}}
\end{figure}

GRB 100414A triggered the \Suzaku WAM on 2010 April14
02:20:22.879(=T0). The WAM \citep{yamaoka09} is a lateral shield of
the Hard X-ray Detector \citep{hxd} on board the \Suzaku
satellite \citep{suzaku} and can function as a GRB monitor sensitive
to the 50$-$5000 keV gamma-rays (e.g. \citet{onda, ohno08, tashiro}).
This burst was detected by all four of the WAM shields, and its
light curve is shown with 1 s time resolution in Figure
\ref{wamlc}. The duration $T_{\rm 90}$ was $21.2\pm0.1$ s which
places it in the class of long duration events.

We also performed the spectral analysis using the WAM transient data
with 55 energy channels and 1-s time resolution. We used the data
from the WAM2 detector only because this particular GRB was situated on
a direction along the normal vector of the WAM2.  The WAM2
spectrum was extracted by integrating over T0$-$1.5 s to T0+24.5 s
with the {\tt hxdmkwamspec} in the standard HEADAS 6.9 package
provided by NASA/GSFC.  The background spectrum was estimated by
incorporating the best-fit model of source-free time region on both
sides (T0$-$504.5 to T0$-$4.5 and T0+37.5 to T0+237.5) of the burst time
region. The energy response was calculated by the WAM response
generator version 1.9 \citep{ohno06}.

We performed the spectral fitting in Xspec version 12.5 and found
that the best-fit model was the GRB Band function
($\chi^2/{\rm dof} = 24.4/24$).  Two other models we also attempted were a
power-law ($\chi^2/{\rm dof} = 253.5/26$) and a power-law with an
exponential cutoff ($\chi^2/{\rm dof} =35.3/25$).  The best-fit parameters
for the Band function were $\alpha$=--0.62(--0.15, +0.17),
$\beta$=--3.05(--0.28,+0.20), and $E_{\rm peak}$=616$\pm$34 keV.  Thus
obtained, the fluence in the 100-5000 keV range was
(6.65$\pm$0.21)$\times$10$^{-5}$ erg cm$^{-2}$ s$^{-1}$.

The corresponding {\it Fermi}/LAT photon data were downloaded from the
Fermi Science Support Center. Based on the likelihood and aperture
photometry\footnotemark, we generated the light curve for the $>100$
MeV energy range (Figure \ref{wamlc}). The highest energy photon from
this event was $\sim4.3$ GeV at approximately 40 s after the burst.
This result was consistent with \citet{lat}.  We also confirmed the
temporal extended emissions.
\footnotetext{http://fermi.gsfc.nasa.gov/ssc/data/analysis/scitools/}

\subsection{Optical follow-up}

We performed the optical afterglow observations using the Canada
France Hawaii Telescope (CFHT)/MegaCam within the framework of the
EAFON \citep{eafon}. The $r'$-band monitoring observations covered the
duration from 5.2 days to 57.6 days after the burst. Additional $g'$-
and $i'$-band images were taken on the night of 2010 April 20 (6.2
days after the burst). Because of the bright moon phase, there was no
optical observation between 7 and 26 days.  The standard CFHT's data
pipeline reduced all of the data.

Figure \ref{optlc} shows the $g'$-, $r'$- and $i'$-band light curves with
several GCN points \citep{uvot,redshift,grond}.  The light curves
demonstrated the steep decay occurring between 3 and 7 days after the
burst. The temporal indices were $-2.3$, $-2.6\pm0.1$ and $-2.7$ in
$g'$-, $r'$-, and $i'$-bands, respectively.  As shown in Figure \ref{optlc} the
spectral energy distribution (SED) of optical afterglow during the steep decay phase (at $\sim$6.2
days) could be satisfactorily expressed according to a simple
power law with the spectral index $\beta = 1.2\pm0.2$.

After the 35 days, a possible host galaxy component dominated.  Because
there was no significant brightness change between 35.2 and 57.6 days,
we estimated the brightness of the host to be $r'=24.7\pm0.1$ AB mag.
The afterglow at 27.2 days exhibited clear excess from the brightness
of the host galaxy. The excess was also fairly inconsistent with the
extrapolation of the afterglow component.

\section{Discussions}
\subsection{Optical afterglow and Jet break}

The prior optical afterglow of GRB100414A ($t<10$ days) exhibited
power-law decays with temporal break at approximately $\sim 2.3$ days.
Despite the paucity of observations for the optical afterglow, the
joint light curve using \Swift/UVOT strongly suggested a temporal break.
The decay indices before and after the break were $\sim 1$ and $\sim
2$, respectively, fully consistent with typical well-observed long GRB
optical afterglows.
Comparing the sample (listed in \citep{mcbreen10}) revealed that the
optical afterglow brightness at 2 days exhibited average brightness or
was rather brighter than those of long GRBs afterglows.

To examine whether the jet model is applicable to present
afterglow, we utilized the relations by \citet{sari99} and calculated
$p$ and $\alpha$ based on the observed $\beta$.  The observed values
were in agreement with the calculated value of $\alpha=2.4\pm0.4$,
assuming that we were observing a jet expansion phase in the frequency
range above the synchrotron cooling.  In addition, although the X-ray
afterglow observation was also poor, the decay index $\alpha_{\rm
  X}=2.3\pm0.4$ between 2.0 and 7.4 days was also consistent with the
jet model. 
The case of wind density profile with $\nu_{m}<\nu_{\rm opt}<\nu_{X}<\nu_{c}$ also
satisfied the closure relation. However, it is unlikely for the late ($6.2$
days) afterglow.

We also attempted to fit the $g'$-band light curve including the UVOT data
using the broken power-law model by fixing the temporal decay indices of
former and post jet break at $-1.3$ and $-2.6$, respectively. The UVOT
data were calibrated according to the SDSS field stars that have similar colors
to the GRB afterglow, and were converted to $g'$ band magnitudes. As shown
in Figure \ref{optlc}, we successfully fitted the light curve and
obtained the jet break time as $t=2.3\pm0.2$ days.  (Figure
\ref{optlc}).
The jet opening angle derived from this temporal break was $\sim
5^{\circ}.8$, consistent with those of other well-observed long GRBs.


%

\subsection{Energetic and Correlation}

The abundance of the multi-wavelength data for estimating the jet
break time makes GRB100414A one of the most favorable targets for evaluating the
${E^{\rm src}_{\rm peak}-E_{\rm iso}}$ and ${E^{\rm src}_{\rm
    peak}-E_{\rm \gamma}}$ relations of studies that had stagnated
because of a lack of data on the ${E^{\rm src}_{\rm peak}}$ estimation
and long-term optical monitoring. In particular, the current event was
the first \fermi/LAT detected event exhibiting the clear jet break
features.

To examine the correlations with the least systematics
errors, we collected five LAT-detected events observed using the {\it
  Suzaku}/WAM, as summarized in Table \ref{tbl-2}.  The spectral
analysis of the WAM data was performed in the same manner as
described in Section 2.1.  
The estimated ${E_{\rm iso}}$ of GRB 090902B in the 1 keV to 10 MeV
range using {\it Suzaku}/WAM are consistent with the value derived
from the \fermi/GBM observation \citep{cenko}).  In the case of
GRB090926A, the value is slightly smaller than that of the estimation
(2.3$\times10^{54}$ erg) using \fermi/GBM \citep{mcbreen10}. The most
likely reason for this inconsistency is that the energy range of
\Suzaku/WAM is insufficient in detecting the underlying power-law
component in the \fermi/LAT-GBM spectrum\citep{090902b}.
GRB090510 was the short event, and the spectral parameters were not
effectively constrained, we excluded this event for the evaluation of correlation. 
The marginal optical temporal break time $t_{\rm jet}$ of GRB090926A
and GRB090902B was reported by \citet{cenko}, \citet{mcbreen10},
\citet{rau} and \citet{swenson}.  As summarized in Table \ref{tbl-2},
the values vary widely according to the estimation methods.
We also excluded the GRB091003 for the evaluation of the ${E^{\rm
    src}_{\rm peak}-E_{\rm \gamma}}$ relation because of a lack of the
${t_{\rm jet}}$ estimation. Interestingly, the three selected events
(GRB090926A, GRB090902B and the current one) shared commonality in the
delayed highest energy photons from their main burst (GRB090926A;
Ackermann et al. 2011, GRB090902B; Abdo et al. 2009). 
 
\begin{figure}
\epsscale{.80}
\plotone{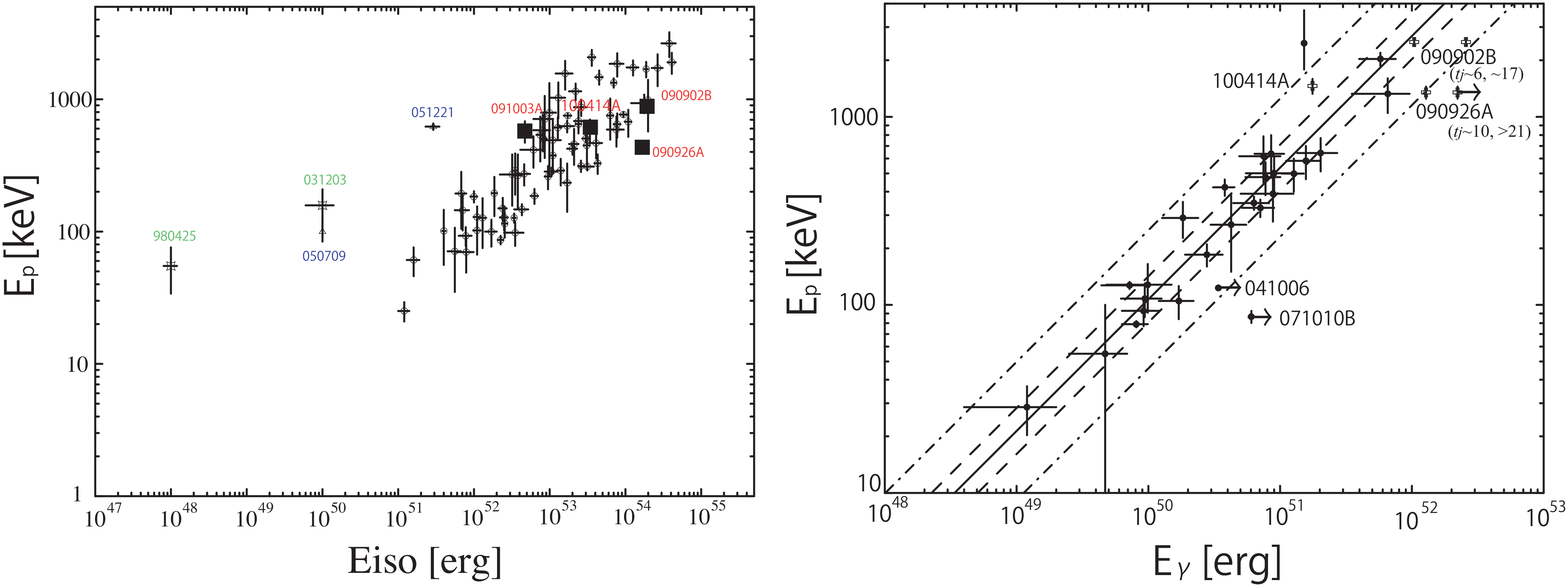}
\caption{Left: The $E^{\rm src}_{\rm peak} - E_{\rm iso}$ relation of \citet{amati} combined with the data points of GRB100414A, GRB091003A, GRB090926A and GRB090902B (filled square). Short duration GRBs and sub-energetic events are indicated by an open square and a triangle, respectively. Right: The $E^{\rm src}_{\rm peak}-E_{\rm \gamma}$ relation including the data for GRB100414A, GRB090926A and GRB090902B corrected for a homogeneous circumburst medium. For GRB090926A and GRB090902B, there are two points based on the estimation of their jet break time summarized in Table \ref{tbl-1}. The solid line indicates the best fit correlation  derived by \citet{g07}. The dashed and dash-dotted lines indicate  the 1$\sigma$ and 3$\sigma$ scatter of the correlation, respectively.\label{grel}
}
\end{figure}

Figure \ref{grel} shows the ${E^{\rm src}_{\rm peak}-E_{\rm iso}}$
and ${E^{\rm src}_{\rm peak}-E_{\rm \gamma}}$ relations with {\it
  Suzaku}/WAM observed LAT-detected events.  We found that all four
events were highly consistent with previous studies of the ${E^{\rm src}_{\rm peak}-E_{\rm iso}}$.  As shown in Figure \ref{grel},
the measured values of current event (${E^{\rm src}_{\rm peak}} = 1458.7^{+132.6}_{-106.6}$ keV and ${E_{\rm iso} = 34.5^{+2.0}_{-1.8}\times 10^{52}}$ erg) adhered closely to the
${E^{\rm src}_{\rm peak}-E_{\rm iso}}$ relation.  Although the
background GRB physic of this correlation is not yet completely
understood, the characteristics of different sub-classes of GRBs
become apparent (Amati 2006). To compare other categories of the
events (e.g., short duration GRBs and sub-energetic events), we
plotted them in Figure \ref{grel}. These features were clearly
different from those of the current event.  The present event and GRB
090902B were also consistent with ${E^{\rm src}_{\rm peak}-E_{\rm \gamma}}$ correlation (Figure \ref{grel}). In the case of
GRB090926A, $E_{\rm \gamma}$ thoroughly depends on the jet break time
estimation.

\subsection{Origin of the late time optical bump}

The late afterglow indicated a flatter evolution with the significant
excess at $\sim27.2$ days (11.5 days in the rest frame). The flatter
component in the light curve was due to the contribution from the host
galaxy of GRB100414A. The estimated brightness of the host was
consistent with those of typical host galaxies at $z\sim1$
(e.g., \citet{host, host2}).  The most straightforward explanation for the
excess in the late afterglow light curve (approximately 27.2 days) is that GRB
100414 was accompanied by a contemporaneous SN.
Several GRBs have bumps in the optical light curves approximately 1$\sim$5
days in the rest frame (e.g., GRBs 030329, 090323, 090328) that can be
interpreted by refreshed shocks (e.g., \citet{rees, 030329}) or late
central engine activity. However, the appearance time of the current event is
significantly later.
The recent spectroscopic confirmation of the GRB$-$SN connection on the
typical long GRB with the standard optical afterglow light curve while
obeying the ${E^{\rm src}_{\rm peak}-E_{\rm iso}}$ relation
\citep{grbsn} is also one of the most encouraging results supporting
the SN association.  Although spectroscopic confirmations are
required, discussing the photometric evaluation is still vital.  In
particular, in the case of rare \fermi/LAT events, light curve
evaluation is usually sufficient for the first step.
 
\begin{figure}
\epsscale{.80}
\plotone{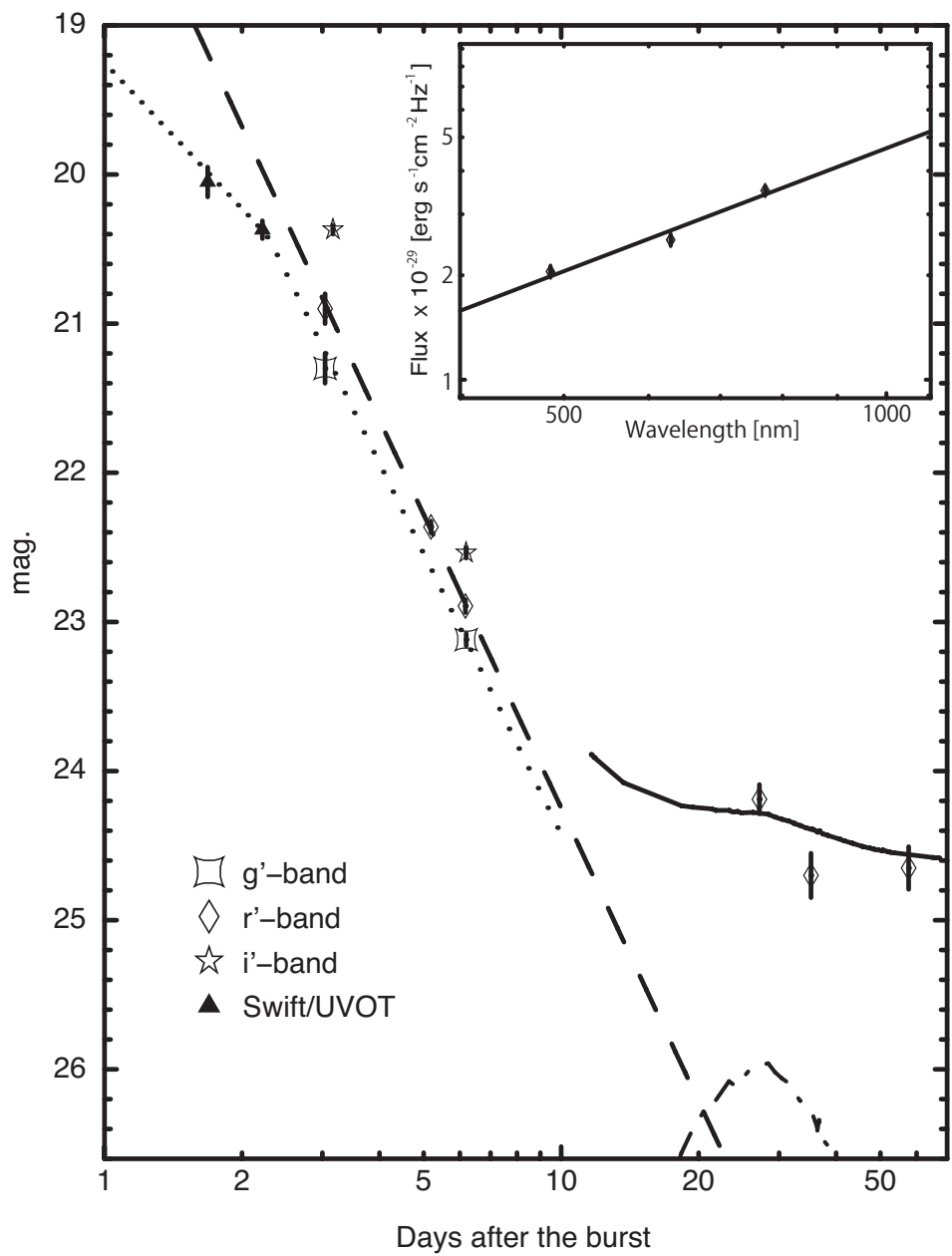}
\caption{Optical light curves and SED. The solid line indicates the model light curve, which is the sum of the afterglow, host galaxy, and supernova component. The dashed and dashed-dotted lines show the best fitted power-law function of the optical afterglow and the template of SN1998bw, as scaled to a redshift of $z=1.368$.  The sub-panel shows the optical afterglow SED at $\sim$6.2 days after the burst.\label{optlc}}
\end{figure}

Several GRBs are associated with SN components, whose rise times and
peak magnitudes are constrained by their optical afterglow light
curves. The GRB$-$SN are globally similar regarding the rise times and
peak magnitudes. The peak times tend to be clustered in 10$-$20
days after the burst in the rest frame, and the absolute magnitudes
are usually approximately $-19$ mag. To examine the consistency of the
possible SN component, we added the light curve template of SN1998bw
to the GRB100414A's host galaxy and afterglow with scaling to the
redshift of $z=1.368$, and fixed the peak time at 27.2 days.  This was
because having less observational data points meant that constraining
the peak time was not possible.  Although the peak time at 27.2 days
(12.2 days in the rest frame) was earlier than those of the GRB$-$SN
events, it was still consistent with the global feature, especially
for the GRB060218A case.  As shown in Figure \ref{optlc}, the combined
light curve was marginally consistent with the observational results.
This implies that the energetic GRB100414A was linked to a core
collapse supernova.

The aforementioned discussions suggest that the current event show no
significant differences with long duration GRBs detected in pre-\Swift
and \Swift eras, despite energetic GeV radiation. By combining the features
of the bumps in the optical light curves and correlations, we
concluded that the progenitor of the current event was likely the
massive stars.

\acknowledgments

This work is supported by grants NSC 98-2112-M-008-003-MY3 (YU) and
99-2112-M-002-002-MY3 (KYH).  Access to the CFHT was made possible by
the Ministry of Education and the National Science Council of Taiwan
as part of the Cosmology and Particle Astrophysics (CosPA) initiative.

\begin{table}
\begin{center}
\caption{Log of CFHT follow-up observations.\label{tbl-1}}
\begin{tabular}{cccccc}
\tableline\tableline
Date       & Start Time(UT) & Delay (days) & Filter & Exposure (s) & mag \\
\tableline
2010-04-19 & 6:36:31.94 & 5.186  & $r'$ & 300 s $\times  5 $ & 22.363 $\pm$ 0.039 \\ 
2010-04-20 & 6:14:39.03 & 6.167  & $r'$ & 300 s $\times  3 $ & 22.894 $\pm$ 0.043 \\ 
2010-04-20 & 6:33:24.63 & 6.188  & $i'$ & 300 s $\times  3 $ & 22.537 $\pm$ 0.035 \\ 
2010-04-20 & 6:52:18.35 & 6.193  & $g'$ & 300 s $\times  3 $ & 23.113 $\pm$ 0.043 \\ 
2010-05-11 & 5:48:04.70 & 27.16  & $r'$ & 300 s $\times  6 $ & 24.189 $\pm$ 0.098 \\ 
2010-05-19 & 7:23:08.96 & 35.239 & $r'$ & 360 s $\times 10 $ & 24.700 $\pm$ 0.149 \\ 
2010-06-11 & 7:26:45.20 & 57.642 & $r'$ & 360 s $\times 10 $ & 24.650 $\pm$ 0.143 \\ 
\tableline
\end{tabular}
\end{center}
\end{table}

\begin{table}
\begin{center}
\caption{{\it Suzaku}/WAM observations of {\it Fermi}/LAT events with redshift.\label{tbl-2}}
\begin{tabular}{ccccccccc}
\tableline\tableline
GRB & $z$ & T$_{90}$ [s] &  E$_{\rm peak}$ [keV] & ${t_{\rm jet}}$ [d] & $E_{\rm iso}$ [10$^{52}$ erg] & $E_{\rm \gamma}$ [10$^{50}$ erg] & $E$h [GeV] \\
\tableline
100414A & 1.368                   & 21   & $616^{+56}_{-45}$  & 2.3$\pm$0.2 & 34.5$^{+2.0}_{-1.8}$    & 17.6 & 4.3 (40s) \\
091003  & 0.8969\tablenotemark{a} & 23   & $576^{+106}_{-72}$ & ---              & 4.7$^{+0.6}_{-0.4}$     & ---  & --- \\
090926A & 2.1062\tablenotemark{b} & 13   & $434^{+32}_{-30}$  & $\sim$10\tablenotemark{e} or $>21$\tablenotemark{f}        & 167.3$^{+11.9}_{-8.4}$  & 128 or $>222$ & 19.6 (25s) \\
090902B & 1.822\tablenotemark{c}  & 19   & $885^{+39}_{-38}$  & $\sim$6\tablenotemark{e} or $\sim$17\tablenotemark{g}  & 193.3$^{+5.9}_{-3.2}$    & 104 or 257  & 33.4 (82s)\\
090510  & 0.903\tablenotemark{d}  & 0.33 & ---               & ---              & ---                   & --- & 30.5 (0.829s)\\
\tableline
\tablenotetext{a}{\citet{091003z}}
\tablenotetext{b}{\citet{090926z}}
\tablenotetext{c}{\citet{090902z}}
\tablenotetext{d}{\citet{090510z,mcbreen10}}
\tablenotetext{e}{\citet{cenko}}
\tablenotetext{f}{\citet{rau}}
\tablenotetext{g}{\citet{mcbreen10}}
\end{tabular}
\end{center}
\end{table}

\end{document}